\documentstyle[12pt]{article}
\newcommand{\be}{\begin{equation}}
\newcommand{\ee}{\end{equation}}
\newcommand{\ba}{\begin{array}}
\newcommand{\ea}{\end{array}}
\newcommand{\beqa}{\begin{eqnarray}}
\newcommand{\eeqa}{\end{eqnarray}}

\newcommand{\noi}{\noindent}

\newcommand{\del}{\partial}

\renewcommand{\ln}{{\rm ln}}

\newcommand{\matr}{\left( \begin{array}}
\newcommand{\ematr}{\end{array} \right)}

\newcommand{\lsim}{\;\raise0.3ex\hbox{$<$\kern-0.75em
\raise-1.1ex\hbox{$\sim$}}\;}
\newcommand{\gsim}{\;\raise0.3ex\hbox{$>$\kern-0.75em
\raise-1.1ex\hbox{$\sim$}}\;}

\begin{document}

\begin{titlepage}

\mbox{}\hfill\makebox[4cm][l]{TURKU-FL-P11}\newline
\mbox{}\hfill\makebox[4cm][l]{HU-TFT-94-15}\newline
\vfill
\Large

\begin{center}
{\bf  Baryon and lepton number transport
     in electroweak phase transition }

\bigskip
\normalsize
{${\rm J. Maalampi}^a, \:\:  {\rm J. Sirkka}^{b}
\:\: {\rm and} \:\: {\rm I. Vilja}^{b}$\\[15pt] $^a$
{\it Department of Theoretical
Physics,
University of Helsinki}\\$^b${\it Department of  Physics,
University of Turku}}

{May 1994}

\bigskip


\vfill

\normalsize

{\bf\normalsize \bf Abstract}

 \end{center}

\normalsize

We consider the baryon number generation by charge transport mechanism
in the electroweak phase transition
taking properly into account thermal fluxes through the wall separating
true  and false vacuum  in the spatial space. We show that the
diffusion from the true vacuum to the false one has a large diminishing
effect on the baryon number unless the wall velocity is near to, but less than,
the speed of sound in the medium  and the ratio between the
collision rate and wall thickness is about 0.3. The maximum net baryon
density generated is $\rho_B/s\simeq 0.2\times 10^{-10}$, where $s$ is
the entropy density of the Universe.
If the wall proceeds as a detonation, no baryon number is produced.

\end{titlepage}

\newpage

\setcounter{page}{2}


It was pointed out by Sakharov \cite{Sakharov} that the cosmological asymmetry
between baryons and antibaryons can follow from the particle interactions in
the early Universe if the following conditions are simultaneously satisfied:
the baryon number $B$ is not conserved, $C$ and $CP$ are not exact symmetries
and the Universe is out of thermal equilibrium. The electroweak
theory has all these ingredients, if the weak phase transition is first order
\cite{Kuzmin}. The baryon number non-conservation follows from the anomalous
violation of the $B+L$ number ($L$ is the lepton number) due to
non-perturbative
sphaleron effects \cite{Manton}. In a first order phase transition there will
be  local departure from thermal equilibrium in the vicinity of the
walls between the symmetric and broken phases. In the standard electroweak
model there also is built in  $CP$ violation manifested as a complex phase in
the Kobayashi-Maskawa quark mixing matrix. The resulting $CP$ violating
effect is, however, at least two orders of magnitude too small to account for
the observed baryon to entropy ratio $\simeq 10^{-10}$. One thus have to go
beyond the minimal standard model.

One possible mechanism, so-called charge transport mechanism, that may produce
a large enough baryon asymmetry was suggested by Cohen, Kaplan and Nelson
\cite{CKN}. The idea is that a net charge, e.g. lepton number, is reflected
from the domain wall between broken and unbroken phase, and anomalous SU(2)
effects in the unbroken phase transmute this into a net baryon number.
Then the baryons produced
pass thermally to the broken
phase where they are stable. In order  to produce a net
charge, the reflected particles have to undergo $CP$ violating interactions
in the wall. In \cite{CKN} these are the Yukawa interactions of neutrinos which
may involve unremovable phases. Such situation may be realized, for example,
in the context of the singlet Majoron model \cite{CMP} and the  left-right
symmetric model \cite{Pati}.

In this paper we shall reconsider some aspects of the charge transport
mechanism. We derive coupled Boltzmann equations for the evolution of the
baryon and lepton numbers in the broken and unbroken phases.
In particular, we shall take into account also the
diffusion of particles from one phase to another, and we actually  find out
that the diffusion  has in many cases quite large diminishing effect on the
creation of the  baryon number.

We are considering a system of an expanding bubble of the true vacuum,
i.e. of the broken phase. The expansion starts with a critical bubble  which
is a spherical configuration of the order parameter \cite{CGM}.  We assume
that the bubble remains spherical during the expansion, its radius being
$R(t)=R_0+v_wt$, where $v_w$  is the velocity of the moving wall and $R_0$ is
the radius of the critical bubble.  The time needed for completing the phase
transition depends on the size of the critical bubble. In the following we
will assume a weak phase transition in which case the critical bubble is large
and the growth time of the bubble is of the same order than the bubble size.
The growth time is given by \cite{EIKR}

\be
t_{growth} = {0.073\over
\sqrt{g_*}}{M_{Pl}\over T_c^2} \left[ \ln {M_{Pl}^4\over T_c^4} \right
]^{-3/2},
\ee
where $g_*$ is the effective number of degrees of freedom (here $g_* \simeq
110$), $M_{Pl}$ is the Planck mass and $T_c$ is the
critical temperature. Using the result of \cite{EIKR} one finds out that the
 bubble radius reaches the value

\be
R_1 = (8\pi)^{1/3} v_w t_{growth} \simeq
3.2\times 10^9 {v_w/{\rm GeV}}
\label{R1}\ee
 upon completion of the phase transition. The
critical bubble radius is thus

\be R_0 = R_1 - v_wt_{growth} \simeq 2.1\times 10^9
{v_w/{\rm GeV}}.
\label{R0}\ee

Strictly speaking, the results above are only for the Standard Model,
but they are likely to give the correct order of magnitudes for a wide range of
models not too different from the  Standard Model, like the Majoron model.
Moreover, our final results are quite insensitive to the precise
values of $R_0$ and $R_1$.
It turns out that the changes on $R_0$ and ¤$R_1$ between as large values as
$10^4 v_w/T$ and $10^{11} v_w/T$ do
not effect the value of the generated baryon
number but by a factor 2, at most.

In the case of a weak first order phase
transition nucleated bubbles grow as deflagrations.
The advancing phase boundary is preceded by a
supersonic shock front which hits the fluid
in the false vacuum and makes it to move to the same direction as  the
boundary.
  The shock front discontinuity is formed within a time period of
$t_0=O(10)\,1/T_c$ which is  very short  compared with
$t_{growth}$ \cite{IKKL}. The velocity of the shock front can be estimated as
\cite{shockbook}

\be  v_{shock}=\frac 1{\sqrt{3}}
\sqrt{\frac{3T_q^4+T_f^4}{3T_f^4+T_q^4}} \approx \frac 1{\sqrt{3}},
\ee

\noindent
since the difference between the temperature of the symmetric phase
$T_f$ and the temperature in the region between phase and shock fronts
$T_q$ is estimated to be smaller than one percent \cite{IKKL}.
Hence the shock front moves essentially with the speed of sound.
In front of the shock wall the particles are at rest in average; there
is no net flux in this region. Inside the shock wave  particles are
moving with a slow velocity outwards
causing a uniform particle flux. Behind the wall
in the broken phase particles are again at rest in average.

Let us now derive the evolution equations for the lepton numbers $L_1$ and
$L_2$, and for the baryon numbers $B_1$ and $B_2$, where the subscript $1$
refers to the symmetric region and the subscript $2$ to the broken one. We
start by writing the equation of continuity for the lepton and baryon number
densities, $l_i$ and $b_i$, and the radial
fluxes, $\lambda_i$ and $\beta_i$,  in the symmetric
and broken regions. Assuming spherical symmetry the equations read
\beqa
\del_t l_1+\frac 1{r^2}\frac{\del}{\del r}(r^2\lambda_1) & = & -\frac
1{2\tau_B}[(1-x)b_1+(1+x)l_1], \nonumber \\
\del_t b_1+\frac 1{r^2}\frac{\del}{\del r}(r^2\beta_1) & = & -\frac
1{2\tau_B}[(1-x)b_1+(1+x)b_1], \nonumber \\
\del_t l_2+\frac 1{r^2}\frac{\del}{\del r}(r^2\lambda_2) & = & 0, \nonumber \\
\del_t b_2+\frac 1{r^2}\frac{\del}{\del r}(r^2\beta_2) & = & 0.
\label{conequ}\eeqa
 The source terms in the symmetric region are
due to anomalous $B+L$-violating processes, characterized by a time scale
$\tau_B\simeq 1 {\rm GeV}^{-1}$ \cite{CKN}. The parameter $x$
is defined as the equilibrium
value of the ratio $(B+L)/(B-L)$. In the broken region the
source terms can be neglected,
since there the $B+L$ violation effects are expected
to be small \footnote{We assume that the
sphaleron energy is large enough \cite{EKV} to prevent
the baryon number dilution in the
broken phase.}.

We now integrate the first two equations of  (\ref{conequ})
over the volume between the
phase boundary with a radius $R(t)$ and the
shock front with a radius $R_s(t)=R_{s0}+v_st$.
Knowing the exact value of the initial shock front radius $R_{s0}$ is not that
important as the shock wave is formed, as mentioned, in a very short time
period  in comparison with the  whole span of the time evolution.

    The ensuing
equation for the lepton number $L_1$ in the symmetric phase reads ($\dot{L}$
stands for time derivative)

 \beqa \dot{L}_1=4\pi R_s^2l_1(R_s,t)v_s&-&4\pi
R^2(t)l_1(R,t)v_w+ 4\pi R^2(t)\lambda_1(R,t)\nonumber\\
&-&\frac
1{2\tau_B}[(1-x)B_1+(1+x)L_1].
\label{Lequ1}
\eeqa
An analogous equation can be derived for the baryon number $B_1$.  In deriving
Eq. (\ref{Lequ1}) it was assumed that the lepton number flux at the shock front
can be neglected.

{}From the latter two equations of
(\ref{conequ}) one obtains similarly
the time evolution equation for the lepton number  in the broken phase:

 \be
\dot{L}_2=4\pi R^2l_2(R,t)v_w-4\pi R^2\lambda_2(R,t),
\label{Lequ2}
\ee
and an analogous equation for the $B_2$. In deriving the
equations for $L_2$ and $B_2$ it was assumed that the flux
of the lepton number  is
negligible at $r=R_0$ for all times.
Strictly speaking this is not true but serves as a
good approximation.

In order to solve  Eqs. (\ref{Lequ1}) and (\ref{Lequ2}) and the
corresponding equations for the baryon number, one has to know the
number densities and the fluxes of leptons and baryons at  the boundary
of the two phases. There are different effects one should take into
account. The CP violating scattering of neutrinos in the advancing wall
produce a net lepton number flux on the  both sides of the wall. This
effect was inspected in detail in \cite{CKN}. They found numerically
that the thermally averaged net lepton number flux $f_L$, in units of
entropy density, reflected from the wall to the symmetric phase is

\be
- f_L = 2.0\times 10^{-10} - 4.2\times 10^{-12}
\ee
at $T = 200$ GeV, depending on the wall velocity and thickness.

 There is also a flux
of leptons and baryons due to thermal transport of particles from one
phase to another. The driving force of the thermal flux is the particle density
differences over the phase boundary.
The thermal  flux of the lepton number in the symmetric phase, to be denoted by
$\lambda_1$, can be presented in the form  $\lambda_1=-D\nabla l_1$. Here
$\nabla l_1$ is the radial lepton number gradient and
  $D$ is  the  diffusion coefficient.
A simple estimate for the diffusion coefficient
is  $D=\langle v\rangle^2\tau/3$, where
$\tau$ is the average time between particle
collisions \cite{Reichl}. Because all particles in the unbroken phase are
massless and they are still relativistic in the broken phase (except the heavy
right--handed neutrinos) the average particle velocity $\langle v \rangle$ is
about unity. The collision time $\tau$ can be calculated the model given. In
the Standard Model one can estimate $\tau\simeq 0.1 {\rm GeV}^{-1}$.

If the wall between the broken and unbroken phase were
at rest we could approximate thermal flux at the boundary as
$D(l_1(R(t),t)-l_2(R(t),t))/\delta$, where $\delta$ is the thickness of the
wall. The wall is, however, moving with velocity $v_w$. Taking that into
account
the formula  modifies to

\be \lambda_1=\frac
D{2\delta}[l_1(R(t),t)(1+v_w)-l_2(R(t),t)(1-v_w)].
\ee
The factors $(1+v_w)/2$ and  $(1-v_w)/2$ give the fractions of light particles
which are caught by the wall in the symmetric region or catch the wall in the
broken region, respectively.
 Those low-energy leptons, which reflect back to the symmetric region  because
of becoming too heavy  to penetrate  the wall, produce a flux
$2r_\nu Dl_1(R(t),t)(1+v_w)/(2\delta)$.
Here the quantity $r_\nu$ gives the fraction
of the  heavy neutrinos that are reflected, and the overall factor of two is
there due to the fact that in the reflection  the lepton
number changes by two units. In our numerical
calculations we will use $r_{\nu}=1/4$
corresponding to the situation in the
Majoron model \cite{CKN}. Note that there is no
contribution of heavy leptons coming from the broken
region to the symmetric region,
because they are, in average, much slower than the wall.
The thermal fluxes for the
baryon number obtained similarly. We have
approximated  the collision time $\tau$ to be the same for the both cases.

About the net lepton and baryon number densities
we assume a uniform distribution inside
the expanded bubble and a linearly decreasing densities inside the
shock front, i.e. between the radii $R(t)$ and $R_0$,
so that $l_1(R_s)=b_1(R_s)=0$.
This yields

\beqa
L_1 & = & \frac{\pi}3 l_1(R,t)\,(R_s-R)(R_s^2+2RR_s+3R^2),
\nonumber \\
 L_2 & = & \frac{4\pi}3 l_2(R,t)\,(R^3-R_0^3). \label{denequ}
\eeqa

Using these approximations  the evolution of the
lepton and baryon numbers is described by the following set of equations:

\beqa
\frac{1}{3u^2}\frac{dL_1}{du} & = & 1-
\frac{\tau}{6\delta v_w}
\left[(1+v_w)(1+r_\nu)\alpha_1(u)L_1-(1-v_w)(1-r_\nu)\alpha_2(u)L_2\right]
-\alpha_1L_1 \nonumber \\
        &   & -\frac{R_0}{2v_w\tau_B}\left[(1-x)B_1+(1+x)L_1\right],
 \nonumber \\
\frac{1}{3u^2}\frac{dB_1}{du} & = &
-\frac{\tau}{6\delta v_w}
\left[(1+v_w)\alpha_1(u)B_1-(1-v_w)\alpha_2(u)B_2\right]
-\alpha_1B_1 \nonumber \\
        &   & -\frac{R_0}{2v_w\tau_B}\left[(1-x)B_1+(1+x)L_1\right],
 \nonumber \\
\frac{1}{3u^2}\frac{dL_2}{du} & = & -1+
\frac{\tau}{6\delta v_w}
\left[(1+v_w)(\alpha_1(u)L_1-(1-v_w)\alpha_2(u)L_2\right](1-r_\nu)
+\alpha_2L_2, \nonumber \\
\frac{1}{3u^2}\frac{dB_2}{du} & = &
\frac{\tau}{6\delta v_w}
\left[(1+v_w)(\alpha_1(u)B_1-(1-v_w)\alpha_2(u)B_2\right]
+\alpha_2B_2.
\label{finequ}
\eeqa
The lepton numbers $L_i$ and baryon numbers $B_i$, $i=1,2$,
are presented in units of $f_L$
and they are scaled with a reference volume
$V_0=4\pi/3R_0^3$. We have defined a dimensionless
variable $u=R(t)/R_0$ and the functions

\beqa
\alpha_1(u)&=&4[(\bar{u}-u)(\bar{u}^2+2u\bar{u}+3u^2)]^{-1},\nonumber\\
\alpha_2(u)&=&1/(u^3-1),
\eeqa

\noi where $\bar{u}=R_s/R_0=R_{s0}/R_0+v_s/v_w(u-1)$.
The baryon density in the broken
phase, the quantity we are interested in, is then given by

\be
b_2=f_L\frac{R_0^3}{R_1^3}B_2(u=R_1/R_0).
\ee

Before moving to the numerical solutions of our
final equations (\ref{finequ}), we
will make
some general observations. It is quite easy to deduce
from the equations that there are
solutions with no baryon number generation.
This depends on the values of the unknown
parameters like the wall thickness $\delta$ and
the wall velocity $v_w$, as well as the
diffusion time scale $\tau$. It turns out that the decisive quantity is

\be
Q=\frac{\tau}{6\delta v_w}(1-v_w)(1-r_v).
\ee
This can be seen as follows. First make the ansatz that the solution is "almost
trivial", i.e. $B_1$, $B_2$ and $L_1$ are zero.
Then  $L_2$ of the following form solves (\ref{finequ}):
\be
L_2=\frac 1Q(u^3-1)+C(u^3-1)^{1-Q}.
\label{L2sol}
\ee
When $Q>1$ the initial condition $L_2(u=1)=0$ forces the integration
constant $C$  equal to zero. By substituting the solution (\ref{L2sol}) to
eqs. (\ref{finequ})  one then finds
that, as consistency requires, the derivatives of $L_1$, $B_1$ and $B_2$
vanish.   In order to have baryon number generated one has therefore to
require $Q<1$.  Physically this means that the wall have to be thick
enough or it must advance into the unbroken phase rapidly enough to
avoid dilution of the net baryon number due to particle diffusion
through the wall. Note that this is also the asymptotic solution  for
the Boltzmann equations when $R_1/R_0 \gg 10^9$. Physically that is,
however, not a favoured case.

Let us now present our numerical results. The theoretical estimates for
the values of   the wall velocity $v_w$ and the wall thickness $\delta$
vary with the assumptions made of the strength of the phase transition
and the nature of the bubble expansion.  We keep these quantities as
free parameters. As we have found, the wall thickness in fact  appears
in our expressions always along with the free time
$\tau$, the results depending on their ratio $\tau/\delta$ only.

In Fig. 1 we present the values of the generated baryon density
$b_2$ as a function of $v_w$ and $\tau/\delta$. The maximum
value of $b_2$ achieved inside the $b_2=0.2$ contour is about 0.23, and
it corresponds to the wall velocity close to the velocity of sound and
the wall thickness close to, but slightly larger than, the neutrino
free path. We have plotted also the $Q=1$ contour, to right of which no
baryon number is generated, because there the wall moves so slowly or
is so thin that the particle diffusion through the wall dilutes the
baryon number. On the other hand, the decreasing of the net baryon
number with decreasing $\tau/\delta$ follows from the fact that in the
case of a thick wall less baryon number moves into the broken phase
from the unbroken phase where it is created. This decrease is of course
quite independent on the wall velocity as the baryon number flux goes
towards the advancing wall.

In Fig. 1 we have integrated the evolution equations from the initial
bubble radius $R_0$ to the final radius $R_1$, whose values were given
in eqs. (\ref{R0}) and (\ref{R1}). As mentioned, this corresponds to a
weak phase transition where the initial bubbles are quite large and the
expansion time small. We have also investigated the sensitivity of our
result on this assumption. This may be relevant since it is also
possible that the initial bubbles are very small and that the bubble
radius increases several orders of magnitude upon the completion of
the phase transition \cite{IKKL}.  In Fig. 2 we present the generated
baryon number as a function of the ratio $R_1/R_0$ with $R_1$ fixed to
the  value (\ref{R1}). One can conclude from this figure
 that the net baryon number
does not depend significantly on the initial size of the bubble in the physical
region of the ratio $R_1/R_0$.

We have presented our results in terms of the reflected lepton number flux
 $f_L$. This quantity has been estimated in \cite{CKN} for some values
of the parameters. For $\delta \simeq 0.1$ GeV$^{-1}$, $v_w= 0.87$ and the
critical temperature  $T_c=200 $GeV they give $f_L=-1.2\times 10^{-10}s$,
where $s$ is the present entropy density in the Universe. Applying this value
to our results would yield the maximum
net baryon number density $\rho_B/s\simeq
0.24\times 10^{-10}$. This is quite close  but slightly below the allowed
range
$\rho_B/s= (0.4 - 1.0)\times 10^{-10}$ obtained from  the standard
nucleosynthesis \cite{Kolb}. The maximum value of $\rho_B/s$
corresponds the values $\tau/\delta \simeq 0.3$ and $v_w = 0.55$. It is
relatively sensitive to the actual values of wall velocity and ratio
$\tau/\delta$. The present theoretical estimates for these are $\delta
\simeq (10 - 40)/T_c$ and $v_w$ lies between 0.2 and 0.8 \cite{EIKR}.
Note, however,
that if the bubble wall proceeds as a  detonation, i.e. $v_w > v_s$, no
baryon number is created. In that case there is no region preceeding the
bubble wall where lepton number conversion to the baryon number could
take place.

Our result seems to give some indication that the baryon number production
by charge  transport mechanism may not be powerful enough to create the
observed baryon number. Because there is some uncertainty in the actual
value of $f_L$,  the baryogenesis via   this scenario is not
completely ruled out. The value of $f_L$ has been calculated only  for
two wall velocities and for a few wall thicknesses using critical
temperature $T_c = 200$ GeV. Also the baryon number production might be
enhanced if there existed  relatively large initial baryon and/or lepton
number density created before  electroweak phase transition. Such could
be created in the Majoron model by decays of of right--handed neutrinos
\cite{IV} or, in e.g. left--right symmetric model, by some  mechanism
during the symmetry breaking of the right--handed sector. These issues
are currently under study \cite{W}.

\newpage {\bf FIGURE CAPTION}

\noindent {\bf Figure 1.} The net generated baryon density in the
$(\log(\tau/\delta),v_w)$ -plane in units of $-f_L$. With the parameter
values corresponding to the area below the $Q=1$
curve no baryon number is generated. $v_s$ is
the sound velocity.

\noindent {\bf Figure 2.} Dependence of the generated baryon density
on the ratio $R_1/R_0$ with $\tau/\delta=0.2$, $v_w=0.5$ and
$R_1$ fixed to the value (\ref{R1}).
\end{document}